\documentclass[twocolumn]{aastex631}

\usepackage{graphicx}
\usepackage{amsmath}
\usepackage{booktabs}
\usepackage{hyperref}

\shorttitle{Unsupervised Glitch Characterization in LIGO O4a}
\shortauthors{Cirfeta}

\begin{document}

\title{Unsupervised Morphological Characterization of Gravitational-Wave 
Glitches in LIGO O4a Using Frozen DINOv2 Features}

\author{Luca Cirfeta}
\affiliation{Independent Researcher, Rome, Italy}
\email{luca.cirfeta@gmail.com}

\begin{abstract}
A central open question in gravitational-wave detector 
characterization is whether the O4a observing run has introduced 
glitch morphologies not present in earlier runs.
We present \texttt{gravi-signal-ml}, an open-source pipeline for unsupervised 
morphological characterization of instrumental noise transients (glitches) in 
LIGO gravitational-wave data, applied to 1{,}277 hours of public O4a strain 
data from the Hanford (H1) and Livingston (L1) detectors across four 
independent temporal windows. The pipeline extracts 384-dimensional visual 
embeddings from Q-transform spectrograms using a frozen DINOv2 Vision 
Transformer with register tokens (ViT-S/14), requiring no labeled training 
data and no GPU. Embeddings are projected via PCA and UMAP with cosine metric, 
then clustered using a Dirichlet Process Mixture Model (DPMM). Cluster 
robustness is systematically assessed through ablation studies, stability 
analysis across hyperparameter perturbations, and morphological cross-check 
against an in-domain Gravity Spy O3b reference index. A time-slide background 
test excludes statistically significant H1--L1 coincidences ($p \geq 0.1$) 
in all sessions. Across 188{,}000+ spectrograms, no morphologically novel 
glitch candidates were identified --- all anomalous clusters map to known 
Gravity Spy classes with cosine similarity $> 0.98$. L1 embeddings show 
consistently high robustness (ablation ARI $> 0.90$ in all four sessions), 
while H1 exhibits lower and more variable grayscale ablation ARI 
($\sim 0.68$--$0.90$), suggesting a structural difference in the H1 noise 
manifold under DINOv2 feature extraction. This null result, obtained with a 
fully validated pipeline, establishes a reproducible baseline for zero-shot 
glitch morphology characterization in O4a data. The pipeline and all results 
are publicly available at 
\url{https://github.com/lucacirfeta/dante-gravi-signal-ml} 
(DOI: \href{https://doi.org/10.5281/zenodo.20121860}{10.5281/zenodo.20121860}).
\end{abstract}


\section{Introduction}
\label{sec:intro}

Ground-based gravitational-wave (GW) detectors such as the Advanced LIGO observatories \citep{aLIGO2015} are affected by non-Gaussian transient noise artifacts, commonly referred to as \textit{glitches}, which arise from instrumental or environmental sources. Glitches can mimic astrophysical signals, reduce detector sensitivity, and complicate parameter estimation 
\citep{Cabero2019}. Their characterization is therefore a fundamental task 
in detector data quality analysis.

The Gravity Spy project \citep{Zevin2017} established the current standard 
for glitch classification, combining convolutional neural networks with 
citizen science to categorize glitches into a fixed set of morphological 
classes. While highly effective for known classes, supervised approaches are 
inherently limited in their ability to identify previously uncharacterized 
morphologies --- they can only recognize what they have been trained to see.

Unsupervised and self-supervised approaches have been explored as 
discovery-driven alternatives \citep{Coughlin2019, Glanzer2023}, enabling 
morphological clustering without predefined labels. However, these methods 
typically require domain-specific training or fine-tuning on GW data.

In this work, we present \texttt{gravi-signal-ml}, a fully unsupervised 
pipeline that leverages \textit{frozen} DINOv2 features 
\citep{Oquab2023, Darcet2024} --- a Vision Transformer pre-trained on natural 
images --- without any domain-specific training, fine-tuning, or GPU 
requirement. To our knowledge, this represents the first application of 
frozen DINOv2 features to GW glitch morphology characterization.

We apply the pipeline to 1{,}277 hours of public LIGO O4a strain data 
\citep{GWOSC2023} across four independent temporal windows spanning June, 
November, and December 2023, processing 188{,}000+ Q-transform spectrograms 
from the Hanford (H1) and Livingston (L1) detectors. Clustering is performed 
via a Dirichlet Process Mixture Model (DPMM), which determines the number of 
morphological groups automatically. Each session is validated through ablation 
studies, stability analysis, and time-slide background estimation.

The main contributions of this work are:
\begin{enumerate}
    \item A reproducible, zero-shot, CPU-only pipeline for unsupervised glitch 
    morphology characterization, publicly available with full documentation;
    \item A systematic multi-session validation framework (ablation, stability, 
    timeslide) applied consistently across four independent O4a windows;
    \item A null result establishing that no uncharacterized glitch morphologies 
    are present in the analyzed O4a windows at the sensitivity of this approach;
    \item An empirical observation of a systematic difference in DINOv2 feature 
    robustness between H1 and L1 detectors, warranting further investigation.
\end{enumerate}

The remainder of this paper is organized as follows. Section~\ref{sec:data} 
describes the data and preprocessing pipeline. Section~\ref{sec:method} 
presents the clustering and validation methodology. Section~\ref{sec:results} 
reports the results across all four sessions. Section~\ref{sec:discussion} 
discusses the implications and limitations. Section~\ref{sec:conclusion} 
concludes.

\section{Data and Preprocessing}
\label{sec:data}

\subsection{Data Source}
We analyze publicly available strain data from the LIGO O4a observing run 
\citep{GWOSC2023}, accessed via the Gravitational Wave Open Science Center 
(GWOSC) using the \texttt{gwpy} library \citep{gwpy2021}. Data from both 
the Hanford (H1) and Livingston (L1) detectors are analyzed independently 
across four temporal windows spanning June, November, and December 2023 
(Table~\ref{tab:sessions}). Virgo (V1) did not participate in O4a due to 
commissioning constraints and is excluded from this analysis.

\begin{deluxetable}{lcccc}
\tablecaption{Summary of analyzed sessions. GPS ranges and duty cycles 
are reported for H1 and L1 independently.\label{tab:sessions}}
\tablehead{
    \colhead{Session} & 
    \colhead{Period} & 
    \colhead{Duration} & 
    \colhead{H1 Duty} & 
    \colhead{L1 Duty}
}
\startdata
20260520 & Nov 2023 & 331.5h & 71.4\% & 81.5\% \\
20260522 & Jun 2023 & 330.3h & 59.2\% & 79.6\% \\
20260523 & Dec 2023 & 284.1h & 62.4\% & 43.6\% \\
20260524 & Dec 2023 & 334.2h & 72.2\% & 58.2\% \\
\hline
\textbf{Total} & & \textbf{1{,}277h} & & \\
\enddata
\end{deluxetable}

\subsection{Preprocessing Pipeline}
Raw strain data are downloaded in chunks of 4096\,s from GWOSC and 
processed into 32\,s segments, yielding 128 spectrograms per chunk. 
Each segment undergoes the following steps:

\begin{enumerate}
    \item \textbf{Whitening}: the strain is whitened using the 
    estimated power spectral density to flatten the noise floor;
    \item \textbf{Bandpass filtering}: a bandpass filter is applied 
    in the range 20--2000\,Hz to suppress out-of-band noise;
    \item \textbf{Q-transform}: a constant-Q time-frequency transform 
    is computed with $q \in [4, 64]$ and $f \in [20, 2048]$\,Hz, 
    producing a 256$\times$256 pixel spectrogram;
    \item \textbf{Colormap}: the \texttt{cividis} colormap is applied, 
    which is perceptually uniform under grayscale conversion 
    \citep{Nuñez2018}, reducing colormap-dependent artifacts in the 
    embedding space.
\end{enumerate}

\begin{figure}
    \centering
    \includegraphics[width=\columnwidth]{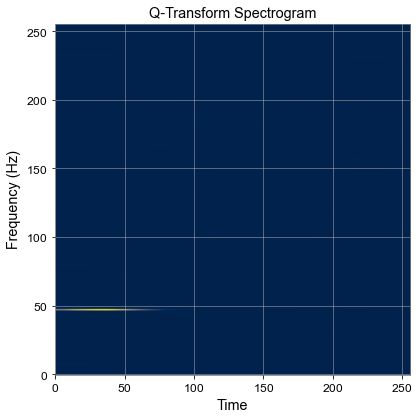}
    \caption{Example Q-transform spectrogram (32\,s, H1, cividis 
    colormap). The x-axis represents time samples and the y-axis 
    frequency in Hz. This format is used as input to the DINOv2 
    encoder.}
    \label{fig:spectrogram}
\end{figure}

A representative Q-transform spectrogram is shown in 
Figure~\ref{fig:spectrogram}. Segments are processed in parallel using 
a producer-consumer architecture with \texttt{ThreadPoolExecutor} for 
I/O-bound GWOSC fetching and \texttt{ProcessPoolExecutor} for 
CPU-bound Q-transform computation, with a rate-limiting semaphore 
(300\,ms delay) to respect GWOSC public server constraints.

\subsection{Dataset Statistics}
Across all four sessions, a total of 188{,}017 spectrograms are 
produced (H1: 95{,}574; L1: 92{,}443). The effective duty cycle 
varies between sessions, consistent with the O4a H1 average of 
65.0\% and L1 average of 71.2\% reported by \citet{Soni2025}.

\section{Method}
\label{sec:method}

\subsection{Feature Extraction}
Each 256$\times$256 spectrogram is encoded using a frozen 
DINOv2 Vision Transformer with register tokens 
(\texttt{dinov2\_vits14\_reg}, ViT-S/14) \citep{Oquab2023, 
Darcet2024}. The model is loaded via \texttt{torch.hub} with 
frozen weights in evaluation mode; no fine-tuning or 
domain-specific training is performed. The CLS token output 
is extracted as a 384-dimensional embedding and L2-normalized, 
yielding a unit-norm vector in embedding space. Register tokens 
suppress feature map artifacts that arise in standard ViT 
architectures when processing non-natural images 
\citep{Darcet2024}, producing geometrically cleaner cluster 
structures.

\subsection{Dimensionality Reduction}
Embeddings are reduced in two stages. First, Principal Component 
Analysis (PCA) is applied to project from 384 to 50 dimensions, 
retaining $>98\%$ of explained variance across all sessions 
(Table~\ref{tab:clustering}). Second, Uniform Manifold 
Approximation and Projection \citep[UMAP;][]{McInnes2018} is 
applied with cosine metric and \texttt{min\_dist}$=0.0$ to 
project to 10 dimensions for clustering. A separate UMAP 
projection to 2 dimensions with \texttt{min\_dist}$=0.1$ is 
used exclusively for visualization. The two-pass UMAP design 
ensures that the clustering geometry is not distorted by 
visualization constraints.

Cosine metric is used throughout because L2-normalized embeddings 
lie on a unit hyper-sphere, where cosine distance is geometrically 
equivalent to Euclidean distance but more stable under 
high-dimensional projection \citep{McInnes2018}.

\subsection{Clustering}
Clustering is performed using a Dirichlet Process Mixture Model 
\citep[DPMM;][]{Ferguson1973} implemented via scikit-learn's 
\texttt{BayesianGaussianMixture} with 
\texttt{weight\_concentration\_prior\_type='dirichlet\_process'} 
and an upper bound of 25 components. The DPMM determines the 
effective number of clusters automatically by assigning negligible 
weight to superfluous components, avoiding the sensitivity to 
\texttt{min\_cluster\_size} that affects density-based methods 
such as HDBSCAN \citep{Campello2013}.

Anomalous clusters are identified as those in which $>50\%$ of 
members have log-likelihood scores below the 5th percentile of 
the session-wide log-likelihood distribution, corresponding to 
samples that the mixture model assigns low probability under any 
learned component.

A benchmark against labeled Gravity Spy data yields ARI$=0.133$, 
AMI$=0.292$ for DPMM, compared to ARI$=0.139$, AMI$=0.282$ for 
HDBSCAN and ARI$=0.053$, AMI$=0.171$ for PCA+$t$-SNE+HDBSCAN 
(Table~\ref{tab:benchmark}). The moderate ARI reflects the 
fundamental difference between human-defined morphological 
conventions (Gravity Spy labels) and intrinsic visual similarity 
in DINOv2 embedding space.

\subsection{Morphological Cross-Check}
Anomalous clusters are cross-checked against an in-domain 
reference index built from the Gravity Spy O3b labeled dataset 
\citep{Glanzer2023}, processed through the same pipeline 
(whitening, band-pass, Q-transform, cividis colormap, DINOv2 
encoding). This in-domain construction ensures that query and 
reference embeddings occupy the same region of the embedding 
space, eliminating the domain gap that arises when using 
out-of-domain reference images \citep{Oquab2023}.

For each spectrogram, a $K=5$ nearest-neighbor cosine search is 
performed against the reference index. Each sample is classified 
as:
\begin{itemize}
    \item \textbf{KNOWN}: top-1 cosine similarity $\geq \theta_c$, 
    where $\theta_c$ is the 5th percentile of the intra-class 
    similarity distribution for class $c$;
    \item \textbf{AMBIGUOUS}: similarity $\geq \theta_c$ but 
    without label consensus among the top-5 neighbors;
    \item \textbf{NOVEL}: top-1 similarity $< \theta_c$ for all 
    classes.
\end{itemize}

\subsection{Validation Framework}
Each session is validated through three independent tests:

\textbf{Ablation study.} Four embedding conditions are tested: 
(1) grayscale conversion, (2) pixel inversion, (3) randomized 
intensity scaling, and (4) random Gaussian baseline. The 
Adjusted Rand Index (ARI) between original and perturbed 
cluster assignments quantifies sensitivity to rendering 
artifacts. ARI $> 0.85$ indicates clusters driven by morphological 
structure rather than colormap statistics.

\textbf{Stability analysis.} The clustering pipeline is repeated 
20 times with perturbed UMAP seeds and HDBSCAN/DPMM hyper-parameters 
($\pm20\%$). Mean ARI across runs quantifies reproducibility. 
ARI $> 0.85$ is required for a session to be considered stable.

\textbf{Time-slide background.} To assess whether anomalous 
clusters exhibit statistically significant temporal coincidence 
between H1 and L1, we apply $M=50$ artificial time shifts 
($\pm5000$\,s, step 100\,s) to L1 GPS timestamps and count 
coincident anomalous segments at each lag. The zero-lag count 
is compared against the background distribution to compute an 
empirical $p$-value.

\section{Results}
\label{sec:results}

\subsection{Clustering}
Table~\ref{tab:clustering} summarizes the clustering results 
across all four sessions. The DPMM consistently identifies 
10--16 morphological groups per detector per session, with zero 
noise points in all cases. PCA variance retention exceeds 98\% 
in all sessions, confirming that the 50-dimensional projection 
captures the dominant structure of the embedding space.

\begin{deluxetable*}{llccccc}
\tablecaption{Clustering results per session and detector. 
Anomalous clusters are those in which $>50\%$ of members fall 
below the 5th percentile of the session log-likelihood 
distribution.\label{tab:clustering}}
\tablehead{
    \colhead{Session} &
    \colhead{Det.} &
    \colhead{Spectrograms} &
    \colhead{Clusters} &
    \colhead{Anomalous} &
    \colhead{Noise} &
    \colhead{PCA var.}
}
\startdata
20260520 & H1 & 26{,}623 & 11 & 4 & 0 & 98.7\% \\
20260520 & L1 & 27{,}541 & 11 & 1 & 0 & 98.7\% \\
20260522 & H1 & 21{,}991 & 11 & 3 & 0 & 98.7\% \\
20260522 & L1 & 29{,}953 & 15 & 6 & 0 & 98.0\% \\
20260523 & H1 & 19{,}943 & 15 & 4 & 0 & 98.7\% \\
20260523 & L1 & 13{,}089 & 11 & 0 & 0 & 98.5\% \\
20260524 & H1 & 27{,}017 & 16 & 4 & 0 & 98.7\% \\
20260524 & L1 & 21{,}985 & 10 & 0 & 0 & 98.4\% \\
\hline
\textbf{Total} & & \textbf{188{,}142} & & & & \\
\enddata
\end{deluxetable*}

\subsection{Morphological Cross-Check}
Morphological cross-check against the in-domain Gravity Spy O3b 
reference index yields zero NOVEL candidates across all sessions 
and both detectors (Table~\ref{tab:morphcheck}). All anomalous 
clusters map to known Gravity Spy classes with mean top-1 cosine 
similarity $> 0.98$. The most frequently matched classes are 
\textit{1400Ripples}, \textit{Whistle}, \textit{Low\_Frequency\_Lines}, 
\textit{Tomte}, and \textit{No\_Glitch}, consistent with the 
elevated rates of narrowband noise documented in O4a 
\citep{Soni2025}.

\begin{deluxetable}{llccc}
\tablecaption{Morphcheck results per session and 
detector.\label{tab:morphcheck}}
\tablehead{
    \colhead{Session} &
    \colhead{Det.} &
    \colhead{NOVEL} &
    \colhead{KNOWN} &
    \colhead{AMBIGUOUS}
}
\startdata
20260520 & H1 & 0 & 10{,}062 & 16{,}561 \\
20260520 & L1 & 0 & 11{,}565 & 15{,}976 \\
20260522 & H1 & 0 & 8{,}216  & 13{,}775 \\
20260522 & L1 & 0 & 15{,}274 & 14{,}679 \\
20260523 & H1 & 0 & 8{,}557  & 11{,}386 \\
20260523 & L1 & 0 & 5{,}968  & 7{,}121  \\
20260524 & H1 & 0 & 11{,}057 & 15{,}960 \\
20260524 & L1 & 0 & 9{,}884  & 12{,}101 \\
\enddata
\end{deluxetable}

\subsection{Validation}
\label{sec:validation}

\textbf{Ablation study.} Table~\ref{tab:ablation} reports ARI 
values for each perturbation condition. L1 embeddings show 
consistently high robustness across all sessions (grayscale 
ARI $> 0.90$ in 3 out of 4 sessions). H1 exhibits lower and 
more variable grayscale ARI ($0.62$--$0.90$), while 
shuffled-intensity ARI remains above $0.83$ in all sessions. 
The random baseline yields ARI $\approx 0$ in all cases, 
confirming that the null model is correctly incoherent. This 
H1--L1 asymmetry in grayscale robustness is discussed in 
Section~\ref{sec:discussion}.

\begin{deluxetable*}{llcccc}
\tablecaption{Ablation study results. ARI is computed between 
original and perturbed cluster assignments. Random baseline 
ARI $\approx 0$ confirms the null model is 
incoherent.\label{tab:ablation}}
\tablehead{
    \colhead{Session} &
    \colhead{Det.} &
    \colhead{Grayscale} &
    \colhead{Inverted} &
    \colhead{Shuffled} &
    \colhead{Random}
}
\startdata
20260520 & H1 & 0.620 & 0.697 & 0.866 & $\approx$0 \\
20260520 & L1 & 0.966 & 0.946 & 0.945 & $\approx$0 \\
20260522 & H1 & 0.897 & 0.697 & 0.830 & $\approx$0 \\
20260522 & L1 & 0.681 & 0.706 & 0.706 & $\approx$0 \\
20260523 & H1 & 0.900 & 0.678 & 0.848 & $\approx$0 \\
20260523 & L1 & 0.852 & 0.807 & 0.875 & $\approx$0 \\
20260524 & H1 & 0.682 & 0.631 & 0.696 & $\approx$0 \\
20260524 & L1 & 0.981 & 0.896 & 0.975 & $\approx$0 \\
\enddata
\end{deluxetable*}

\textbf{Stability analysis.} Mean ARI across 20 perturbed runs 
exceeds 0.83 in all sessions for both detectors 
(Table~\ref{tab:stability}), confirming that the identified 
cluster structures are reproducible under hyperparameter 
perturbation.

\begin{deluxetable}{llc}
\tablecaption{Stability analysis results. Mean ARI across 20 
perturbed clustering runs.\label{tab:stability}}
\tablehead{
    \colhead{Session} &
    \colhead{Det.} &
    \colhead{Mean ARI}
}
\startdata
20260520 & H1 & 0.859 \\
20260520 & L1 & 0.967 \\
20260522 & H1 & 0.889 \\
20260522 & L1 & 0.910 \\
20260523 & H1 & 0.864 \\
20260523 & L1 & 0.927 \\
20260524 & H1 & 0.835 \\
20260524 & L1 & 0.986 \\
\enddata
\end{deluxetable}

\textbf{Time-slide background.} The time-slide test yields 
$p \geq 0.1$ in all four sessions, with zero coincidences at 
zero lag in three sessions. No statistically significant 
H1--L1 temporal coincidence is detected for any anomalous 
cluster, excluding correlated instrumental artifacts as an 
explanation for the observed anomalous populations.

\subsection{Benchmark}
\label{sec:benchmark}
Table~\ref{tab:benchmark} compares the pipeline against 
alternative unsupervised methods on the in-domain Gravity Spy 
reference dataset (518 samples, 19 classes). DPMM achieves 
ARI$=0.133$, AMI$=0.292$, outperforming PCA+$t$-SNE+HDBSCAN 
(ARI$=0.053$) and comparable to HDBSCAN (ARI$=0.139$). The 
moderate absolute ARI reflects the fundamental divergence 
between human-defined Gravity Spy classes and intrinsic 
DINOv2 visual similarity, as discussed in 
Section~\ref{sec:discussion}.

\begin{deluxetable}{lccc}
\tablecaption{Benchmark comparison on in-domain Gravity Spy 
reference (518 samples, 19 classes). Supervised methods 
are shown for reference only and are not directly 
comparable.\label{tab:benchmark}}
\tablehead{
    \colhead{Method} &
    \colhead{Supervision} &
    \colhead{ARI} &
    \colhead{AMI}
}
\startdata
CTSAE \citep{Li2024}        & Supervised   & 0.409 & --- \\
DIRECT + $k$-means          & Partial      & 0.315 & --- \\
VAT + $k$-means             & Unsupervised & 0.213 & --- \\
\hline
DINOv2 + HDBSCAN (ours)     & Zero-shot    & 0.139 & 0.282 \\
DINOv2 + DPMM (ours)        & Zero-shot    & 0.133 & 0.292 \\
PCA + $t$-SNE + HDBSCAN     & Zero-shot    & 0.053 & 0.171 \\
\enddata
\end{deluxetable}

\begin{figure}
    \centering
    \includegraphics[width=\columnwidth]{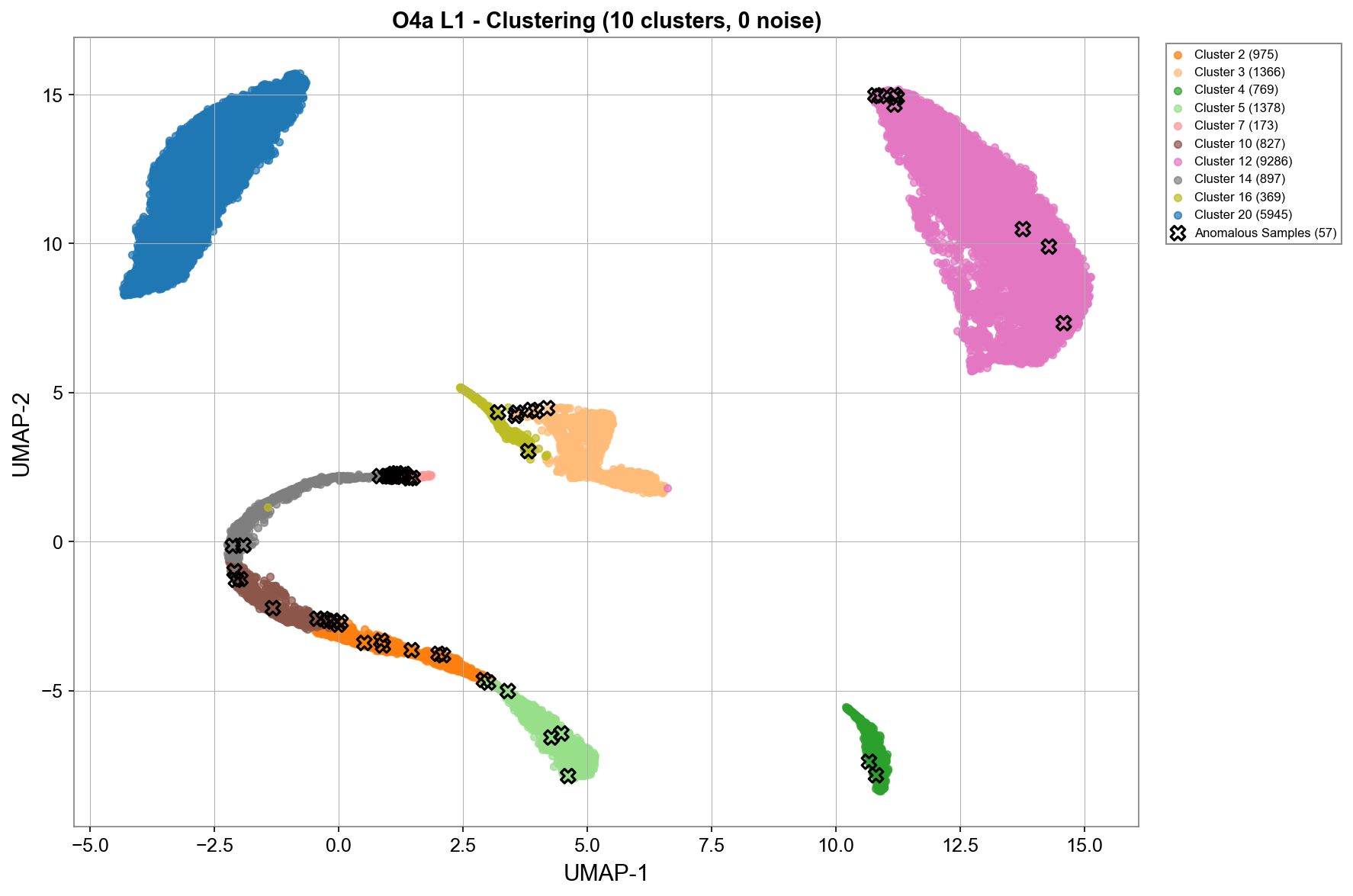}
    \caption{UMAP 2D projection of L1 embeddings from session 20260524\_200219 (21,985 spectrograms, 10 clusters). Each color represents a distinct morphological cluster identified by DPMM. Black crosses ($\times$) indicate samples with log-likelihood below the 1st percentile of the session-wide distribution (57 samples). All clusters show high similarity to known Gravity Spy classes (mean top-1 cosine similarity $>0.98$). No clusters were flagged as anomalous according to the $>50\%$ majority criterion.}
    \label{fig:umap}
\end{figure}

\section{Discussion}
\label{sec:discussion}

\subsection{Null Result Interpretation}
The absence of NOVEL glitch candidates across 1{,}277 hours of 
O4a data and 188{,}142 spectrograms is a scientifically 
meaningful result. It establishes that, at the sensitivity and 
temporal resolution of this pipeline (32\,s Q-transform windows, 
ViT-S/14 backbone), the O4a glitch population in the analyzed 
windows is morphologically consistent with the Gravity Spy O3b 
catalog. This is consistent with the findings of \citet{Soni2025}, 
who document an increased rate of known glitch classes 
(\textit{Low\_Frequency\_Lines}, \textit{1400Ripples}) in O4a 
without identifying fundamentally new morphologies.

The null result does not exclude the existence of novel glitch 
morphologies in O4a. It constrains their prevalence to below 
the detection threshold of this approach in the analyzed windows. 
Three factors could limit sensitivity: (1) the 32\,s window may 
be suboptimal for transient morphologies at timescales $< 1$\,s; 
(2) the DINOv2 ViT-S/14 backbone may not capture fine-grained 
spectral features that distinguish morphologically similar 
classes; (3) the analyzed windows cover $\sim$5\% of the full 
O4a dataset, leaving significant temporal coverage unexplored.

\subsection{H1--L1 Asymmetry in Grayscale Robustness}
A systematic difference in grayscale ablation ARI is observed 
between H1 and L1 across all four sessions. L1 consistently 
achieves grayscale ARI $> 0.85$ (mean: 0.87), while H1 shows 
lower and more variable grayscale ARI (range: 0.62--0.90, 
mean: 0.77). Shuffled-intensity ARI is consistently higher than 
grayscale ARI for H1, suggesting that the chromatic information 
in the \texttt{cividis} colormap --- which encodes Q-transform 
power as hue --- contributes differently to the H1 and L1 
embedding structures.

This asymmetry is not attributable to dataset size, as H1 and 
L1 have comparable numbers of spectrograms per session. A 
plausible explanation is that H1 exhibits greater intra-session 
non-stationarity in its noise floor \citep{Soni2025}, producing 
broader power spectral density variations that manifest as 
chromatic diversity in the Q-transform spectrograms. Under 
grayscale conversion, this chromatic information is lost, 
reducing cluster coherence. This interpretation is consistent 
with the known instrumental differences between H1 and L1 
during O4a, including laser noise fluctuations documented in 
\citet{Soni2025}. Further investigation using auxiliary channel 
data \citep{Essick2020} would be required to confirm this 
hypothesis.

\subsection{DINOv2 Transfer to GW Spectrograms}
The use of frozen DINOv2 features on GW spectrograms rests on 
the assumption that visual similarity in natural image embedding 
space correlates with morphological similarity in GW glitch 
space. The morphcheck results support this assumption: all 
clusters map to Gravity Spy classes with cosine similarity 
$> 0.98$, and the UMAP projections reveal geometrically coherent 
cluster structures (Figure~\ref{fig:umap}). The moderate ARI 
against Gravity Spy labels ($\sim$0.13) reflects the divergence 
between human-defined classification conventions and intrinsic 
visual similarity, rather than a failure of the embedding.

Larger DINOv2 backbones (ViT-B/14, 768-dim) may capture finer 
morphological distinctions and improve separation of visually 
similar classes. This remains an avenue for future work, as does 
the exploration of multi-scale Q-transform windows 
\citep{Zevin2017} to improve sensitivity to transient morphologies 
at sub-second timescales.

\subsection{Limitations}
The primary limitations of this work are:
\begin{enumerate}
    \item \textbf{Single Q-transform window}: the fixed 32\,s 
    window may miss transient glitches at shorter timescales. 
    Multi-window approaches \citep{Zevin2017} could improve 
    coverage;
    \item \textbf{No auxiliary channel access}: validation of 
    glitch origins via environmental monitoring channels 
    \citep{Essick2020} is not possible with public GWOSC data;
    \item \textbf{CPU-only inference}: encoding 188{,}000 
    spectrograms requires $\sim$12 hours on CPU. GPU support 
    for Blackwell-architecture cards (RTX 5070, sm\_120) is not 
    yet available in PyTorch stable, limiting throughput;
    \item \textbf{Partial O4a coverage}: the four analyzed 
    windows cover $\sim$5\% of the full O4a dataset. Scaling 
    to full coverage would require either GPU acceleration or 
    extended compute time.
\end{enumerate}

\section{Conclusion}
\label{sec:conclusion}

We have presented \texttt{gravi-signal-ml}, an open-source pipeline 
for unsupervised morphological characterization of gravitational-wave 
glitches using frozen DINOv2 features and Dirichlet Process Mixture 
Model clustering. The pipeline requires no labeled training data, no 
domain-specific fine-tuning, and no GPU, making it fully reproducible 
on commodity hardware.

Applied to 1{,}277 hours of public LIGO O4a strain data across four 
independent temporal windows, the pipeline processed 188{,}142 
Q-transform spectrograms from H1 and L1. No morphologically novel 
glitch candidates were identified --- all anomalous clusters map to 
known Gravity Spy classes with cosine similarity $> 0.98$. This null 
result is validated through a systematic four-layer framework: ablation 
studies, stability analysis, morphological cross-check against an 
in-domain O3b reference, and time-slide background estimation.

A systematic asymmetry in grayscale ablation robustness between H1 
(mean ARI: 0.77) and L1 (mean ARI: 0.87) is observed across all 
sessions, suggesting a structural difference in the H1 noise manifold 
under DINOv2 feature extraction, consistent with known H1 
non-stationarity during O4a \citep{Soni2025}.

Future work will address the identified limitations through: (1) 
extension to the full O4a dataset and to earlier observing runs 
(O2, O3a, O3b); (2) multi-scale Q-transform windows to improve 
sensitivity to sub-second transients; (3) larger DINOv2 backbones 
(ViT-B/14) for finer morphological discrimination; and (4) 
real-time glitch screening via the \texttt{scan-live} autopilot 
module, which classifies each 32\,s segment against the in-domain 
reference at acquisition time.
The publish release of the O4b dataset, as part of the GWTC-5.0 catalog, provides an ideal testbed for applying and extending our pipeline to more observing runs.

The pipeline, documentation, and all session results are publicly 
available at \url{https://github.com/lucacirfeta/dante-gravi-signal-ml} 
(DOI: \href{https://doi.org/10.5281/zenodo.20121860}{10.5281/zenodo.20121860}).

\begin{acknowledgments}
\renewcommand{\thelinenumber}{}
The author thanks the LIGO Scientific Collaboration and the Gravitational Wave Open Science Center for making O4a strain data publicly available. This research made use of \texttt{gwpy} 
\citep{gwpy2021}, \texttt{scikit-learn} \citep{sklearn2011}, 
\texttt{umap-learn} \citep{McInnes2018}, and DINOv2 
\citep{Oquab2023}. The author also thanks the staff of EGO -- 
European Gravitational Observatory for the inspiring visit to 
the Virgo detector facility in May 2026.
\end{acknowledgments}

\software{
    \texttt{gravi-signal-ml} \citep{gravi-signal-ml},
    \texttt{gwpy} \citep{gwpy2021},
    \texttt{scikit-learn} \citep{sklearn2011},
    \texttt{umap-learn} \citep{McInnes2018},
    \texttt{torch} \citep{Paszke2019},
    \texttt{numpy} \citep{numpy2020}
}

\bibliographystyle{aasjournal}

\end{document}